\documentclass{elsart}
\journal{}

\usepackage[latin1]{inputenc}
\usepackage{longtable}
\usepackage[dvips]{graphicx}
\usepackage{epsfig}
\usepackage{amssymb}
\usepackage{amsfonts}
\usepackage{amsmath}
\usepackage{mathrsfs}
\usepackage{hhline}
\usepackage{array}
\usepackage{graphicx}
\usepackage{dcolumn}
\usepackage{bm}
\usepackage{lscape}

\begin{document}
\begin{frontmatter}

\title{Test of pulse shape analysis 
using single Compton scattering events}
\author{I.~Abt},
\author{A.~Caldwell},
\author{K.~Kr\"oninger$^{**}$},
\author{J.~Liu},
\author{X.~Liu$^*$},
\author{B.~Majorovits}
\address{Max-Planck-Institut f\"ur Physik, M\"unchen, Germany}
\corauth[cor]{email: {\tt xliu@mppmu.mpg.de},
Tel. +49-(0)89-32354-337}
\corauth[cor]{Current address: II. Physikalisches Institut, Universit\"at G\"ottingen, Germany}
\begin{abstract}
Compton scattering is one of the dominant interaction
processes in germanium for photons with an energy of around two MeV.
If a photon scatters only once inside a germanium detector,
the resulting event contains only one electron which normally
deposits its energy within a $mm$ range.
Such events are similar 
to $^{76}$Ge neutrinoless double beta-decay (0$\nu$$\beta\beta$) events 
with just two electrons
in the final state.
Other photon interactions like pair production or
multiple scattering can result in events
composed of separated energy deposits.
One method to identify the
multiple energy deposits is the
use of timing information contained in 
the electrical response of a detector or a segment of a detector.

The procedures developed to separate single- and multiple-site
events~\cite{cite:siegfried_dep_psa} 
are tested with specially selected event samples
provided by
an 18-fold segmented prototype germanium detector for Phase II of
the GERmanium Detector Array, GERDA~\cite{cite:gerda_white_paper}.
The single Compton scattering, i.e. single-site, 
events are tagged by coincidently
detecting the scattered photon with a second detector
positioned at a defined angle.
A neural network is trained to separate such events from
events which come from multi-site dominated samples.
Identification efficiencies of $\approx$~80\% 
are achieved for both single- and multi-site events.
\end{abstract}
\end{frontmatter}


\newpage

\section{Introduction}
\label{section:introduction}

Photons of energies around 2~MeV 
have a high probability to interact in Germanium 
through Compton scattering. The mean free
path of the process is a couple of centimeters. 
If a photon Compton scatters only
once inside a germanium detector, the recoiling
electron deposits its energy most likely within a 1~mm range,
resulting in a so-called single-site event (SSE).
If, in contrast, a photon interacts through pair production
or scatters multiple times,
energy can be deposited at different locations separated 
by typically a few centimeters,
resulting in a so-called multi-site event 
(MSE).
The charge carriers created by the energy 
deposition in the germanium detector
drift towards the anode and cathode of the detector.
While the charge amplitude of the induced pulse
is determined by the number of carriers
(thus by the energy deposited), the pulse shape
is determined by the location(s) of the energy deposition(s)
and thus the charge drifting times.
MSEs are expected to have more involved
pulse shapes than SSEs, and thus, pulse shape analysis (PSA)
can be used to separate the two classes of 
events~\cite{cite:siegfried_dep_psa,cite:first_scs,cite:HdMo,cite:HdMo_dep,cite:igex,cite:dep_part_majorana}.

One application of PSA is the background rejection
in experiments searching for neutrinoless double-beta 
decay (0$\nu$$\beta\beta$) 
in $^{76}$Ge-enriched detectors, 
such as the GERDA experiment~\cite{cite:gerda_white_paper,cite:siegfried_mc}.
The expected 0$\nu$$\beta\beta$ signal events
have two electrons in the final state
with a total energy
of 2.039~MeV.  These are mostly SSEs.
A large fraction of the expected background events
are induced by external photons 
with energy depositions around the Q-value.
These events are expected to be predominantly MSEs
which can be rejected by PSA.

In order to study and improve the performance of PSA,
SSE- and MSE-dominant data samples have to be collected
independently of the pulse shape.
In this paper a method to collect single Compton scattering
events (SCS) as an SSE-dominant sample is
investigated in more detail.
The energy of the scattered photon in an SCS event can be calculated
given the incoming photon energy and the scattering angle.
Therefore, SCS events can be collected
by positioning a second germanium detector
at a specific angle with respect to the first detector
and using it to tag escaped
photons with the correct energy~\cite{cite:first_scs}.
If the incoming photon 
has an energy of 2.614~MeV
as emmitted by  a $^{208}$Tl
source, a photon Compton scattered at 72$\textdegree$
has an energy of 575~keV.
This signature is used to tag 
the single recoiling electron
inside the first germanium detector.
The energy in the event is
equal to the germanium 0$\nu$$\beta\beta$ Q-value.
The location of the energy deposition of the electron
within the detector volume
is controlled by positioning
the source and the second detector correspondingly.

Another common method to collect an SSE-dominant sample
is to select the double-escape
events (DEP)~\cite{cite:siegfried_dep_psa,cite:HdMo,cite:HdMo_dep,cite:dep_part_majorana}.
The incoming photon interacts with the germanium detector
through pair production and
the two 511~keV photons from the positron annihilation
escape the detector without further interaction.
The electron and positron mostly deposit their energies
very locally and result in an SSE.
Another useful sample contains
so-called single-escape
events (SEP) where only one 511~keV photon escapes.
The other photon mostly deposits its energy
at locations different from those of the
electron and positron.
Thus, SEP events provide an MSE-dominant sample
with energy deposition close to the 0$\nu$$\beta\beta$ Q-value.

However, the DEP events are not a perfect test sample for
the expected 0$\nu$$\beta\beta$ events.
If the two photons escape the detector,
the interaction point is more likely 
close to the detector surface as compared to SCS events.
0$\nu$$\beta\beta$ events, on the other hand,
are distributed evenly within the detector volume.
In addition, DEP and 0$\nu$$\beta\beta$ events have different energies.
A DEP event induced by a 2.6~MeV photon from a $^{208}$Tl source
has an energy of 1.59~MeV,
quite different from the 0$\nu$$\beta\beta$ Q-value.
In these respects studies with SCS samples suffer less
from systematic effects.

The experimental setup and the data collection
are described in chapter~\ref{chapter:experiment}.
The Monte Carlo simulation is also included in this chapter.
It is used to verify that the collected SCS samples are SSE-dominated.
In chapter~\ref{chapter:psa} 
a PSA package based on an artificial neural network (ANN) is presented.
The training methods are described and the results given.

\section{Experimental setup, data selection and MC simulation}
\label{chapter:experiment}


\subsection{Experimental setup}
\label{section:setup}

\begin{figure}[th!]
\center
\includegraphics[scale=0.65]{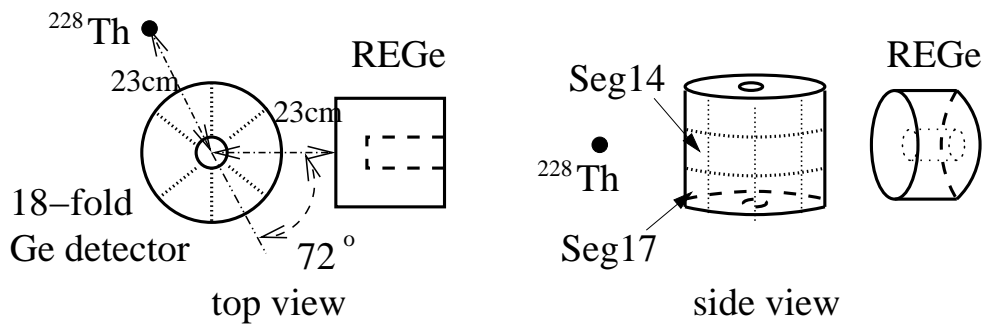}
\caption{Schematic of the experimental setup with the 18-fold segmented
germanium detector as target and the REGe detector to tag photons
at 72$\textdegree$.
The dotted lines illustrate the segment boundaries.} 
\label{fig:setup}
\end{figure}

The experimental setup is 
illustrated in Figure~\ref{fig:setup}.
The segmented germanium detector under study is a prototype detector
for Phase-II of the GERDA experiment~\cite{cite:gerda_white_paper}. 
The true coaxial 18-fold segmented $n$-type HPGe
dectector has a weight of 1.63~kg
and the dimensions are
69.8~mm height and 75.0~mm diameter;
the inner hole has a diameter of 10.0~mm.
The segmentation scheme is 3-fold along
the vertical axis and 6-fold in the azimuthal angle.
(see Figure~\ref{fig:setup}).
Signals from the 18 segments and the core
of the detector are amplified by charge sensitive pre-amplifiers and
read out by a Pixie4 DAQ system~\cite{cite:pixie4_daq}
with 14-bit ADC's at a sampling rate of 75~MHz.
The resolution (FWHM) of the core is $\approx$3.5~keV at 1.3~MeV
and those of the segments are between 2.5 and 4.0~keV.
A time resolution of roughly 10~ns can be achieved with
the sampling rate used. 
This corresponds to a position resolution 
of $\approx$1~mm inside the detector volume.\footnote{
The typical drift velocity of the charge 
carriers inside a germanium detector is 
$\approx$1~cm per 100~ns.}
More information about the segmented detector
and the DAQ system can be found 
in~\cite{cite:siegfried_characterization}.

A 100~kBq $^{228}$Th source is positioned
at a distance of 23$\pm$1~cm from the center of the segmented detector
and faces the center point of segment 14, as illustrated
in Figure~\ref{fig:setup}.
A second non-segmented and well-type germanium detector, 
a Canberra REversed Germanium detector (REGe)~\cite{cite:rege},
is positioned at the same height
with the closed end facing the segmented germanium detector.
The distance from the closed end surface to the center of the
segmented detector is 23$\pm$1~cm.
The REGe crystal is 60~mm in height and 65~mm in diameter.
It has a resolution (FWHM) of 2.3~keV at 1.3~MeV.
It is used to tag the photons scattered mostly in segment~14.
The geometrical acceptance of the
REGe~detector results in recorded SCS events with
scattering angles
between $\approx$~65$\textdegree$ and $\approx$~80$\textdegree$
corresponding to energy depositions in the segmented detector
between $\approx$~1940~keV and $\approx$~2110~keV.
The precision of the alignment of the REGe detector with respect to
the $^{228}$Th source and the segmented detector is $\approx$~5$\textdegree$.

The energy thresholds for all channels are set to 100~keV.
A coincidence trigger is required between the
core of the segmented detector and the REGe
with a coincident time window of 500~ns.
Due to a technical limitation of the coincidence trigger of the DAQ system,
only four channels could be read out.
Thus, for each coincidence trigger,
only the energies of the core ($E_{Core}$), 
segment 14 ($E_{Seg14}$), segment 17 ($E_{Seg17}$)
(below segment 14, as illustrated in Figure~\ref{fig:setup}) 
and the REGe ($E_{REGe}$) were recorded.
300 time samples were taken for each pulse for all 4 channels.
This corresponds to a time window of 4~$\mu$s
including 1~$\mu$s before the arrival of the trigger.
In this analysis, however, only the core pulses are
used for the PSA.

The actual coincidence trigger rate was $\approx$~12~Hz.
The independent trigger rates of the segmented detector and
of the REGe detector were both $\approx$2000~Hz.
This results in an accidental coincidence rate of $\approx$2~Hz.
The coincidence trigger rate 
without the $^{228}$Th source is $<$0.1~Hz.
Therefore, without further cuts,
$\approx$~20\% of all events
are expected to originate from accidental coincidences.\footnote{
The fraction is expected to differ for different energy ranges,
as the trigger rate varies.}
However, the fraction of accidental coincidence events
among the selected SCS events is negligible,
as discussed in the next section.

\subsection{Event selection}

In total 360~000 coincident events were collected.
Four different data samples are selected:

\begin{itemize}

\item $\Gamma_{SCS}$ : Single-Compton-Scattering (SCS) events

\hspace{0.7cm} $|E_{Core}+E_{REGe} - 2614.5 | < $ 5.0~keV

$and$ $1940<E_{Core}<2090$ keV

$and$ $|E_{REGe}-583.2 | > $ 3.0~keV

\item $\Gamma_{2.6}$ ~~: events with the 2.6~MeV 
photon fully absorbed in the segmented detector

\hspace{0.7cm} $|E_{Core} - 2614.5 | < $ 5.0~keV

\item $\Gamma_{DEP}$ : DEP events

\hspace{0.7cm} $|E_{Core} - 1592.5 |< $ 5.0 keV  (Two 511~keV photons escape.)

\item $\Gamma_{SEP}$ : SEP events

\hspace{0.7cm} $|E_{Core} - 2103.5 |< $ 5.0 keV  (One 511~keV photon escapes.)

\end{itemize}

The $\Gamma_{SCS}$~sample is selected through three cuts. 
The allowed window of $\pm$\,5~keV of the sum energy of both detectors 
around 2614.5~keV covers about three times the combined
energy resolution (3$\sigma$) of the detectors.
The geometrical acceptance for SCS~events extends to 2110~keV, but 
SEP~events would contaminate the sample, as they have
a core energy of $E_{Core}$~=~2103.5~keV in this setup.
They are excluded by removing events with the
core energy of the segmented detector above~2090~keV.
The $^{208}$Tl decay also produces 583.2~keV photons with
a branching ratio of 84.5\%. 
To avoid coincidences orginating from
these photons an energy window of $|E_{REGe}-583.2 | <$~3.0~keV
is excluded.

The single-segment events are selected from
each data sample by additionally requiring 
\begin{itemize}

\item single-segment requirement:

$|E_{seg14}-E_{Core}|<5.0$~keV $or$ $|E_{seg17}-E_{Core}|<5.0$~keV

\end{itemize}
The single-segment event samples are noted as
$\Gamma_{SCS}^{S}$, $\Gamma_{2.6}^{S}$, $\Gamma_{DEP}^{S}$ and
$\Gamma_{SEP}^{S}$, respectively.

The coincidence trigger is only relevant for
the $\Gamma_{SCS}$ sample.
However, the other samples are selected
out of the collected coincident events
to ensure the same experimental conditions.
In principle the REGe detector could also 
be used to tag 511~keV photons
for events in the $\Gamma_{SEP}$ and $\Gamma_{DEP}$ samples.
However, the statistics available is not sufficient.

\begin{figure}[th!]
\center
\includegraphics[scale=0.75]{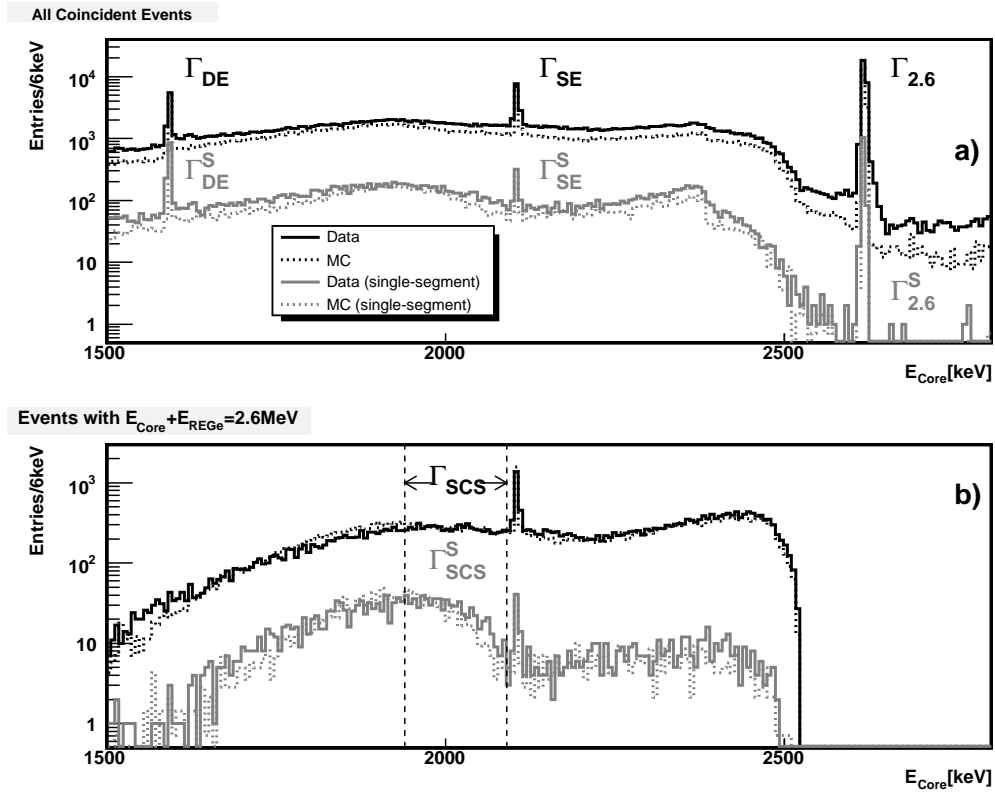}
\caption{$E_{Core}$ distributions a) for all coincident events
b) for events with $E_{Core}$+$E_{REGe}$=(2614~$\pm$~5)~keV.
The 8 selected samples are indicated.
The predicted distributions from the Monte Carlo are shown as well.}
\label{fig:energy_data_mc}
\end{figure}

The distribution of the energy of the core, $E_{Core}$, 
of all coincident events is shown
in Figure~\ref{fig:energy_data_mc}a.
The $E_{Core}$ distribution of all single-segment coincident events 
is shown in the same plot. Also shown are the simulated spectra
which will be discussed in the next section.
Figure~\ref{fig:energy_data_mc}b 
shows the $E_{Core}$ distribution
for all coincident events with 
$E_{Core}+E_{REGe} = (2614.5 \pm 5.0)$~keV.
The arrows indicate the $E_{Core}$ range 
corresponding to the acceptance angles for 
the $\Gamma_{SCS}$ sample.
\begin{table}[th!]
\center
\caption{The numbers of events in all data samples are
presented in the first row.
For the $\Gamma_{SCS}$ sample,
$f_{c}$ in the second row
corresponds to the fraction of events
with $|\Delta T|>$~107~ns.
For the $\Gamma_{2.6}$, $\Gamma_{DEP}$ and $\Gamma_{SEP}$,
it corresponds to the fraction of events in
the central peaks of the $|\Delta T|$ distributions.
The ratios of event numbers for data and MC
are given in the third row
with statistical errors only.}
\begin{tabular}{|c|c|c|c|c||c|c|c|c|}
\hline
sample     &  $\Gamma_{SCS}$ & $\Gamma_{2.6}$
           & $\Gamma_{DEP}$ & $\Gamma_{SEP}$
           &  $\Gamma_{SCS}^{S}$ & $\Gamma_{2.6}^{S}$
           & $\Gamma_{DEP}^{S}$ & $\Gamma_{SEP}^{S}$ \\ \hline
\#events & 6,716 & 25,780 & 6,898 & 10,093
        & 642 & 1,131 & 1,059 & 411 \\ \hline
$f_{c}$ [\%] &   $>$99
&   78$_{\pm1}$
&   87$_{\pm1}$
&   85$_{\pm1}$
&   97$_{\pm4}$
&   78$_{\pm2}$
&   87$_{\pm3}$
&   82$_{\pm4}$
\\ \hline
\# MC/data [\%]  & 103$_{\pm1}$ & 66$_{\pm1}$
           & 80$_{\pm1}$ & 79$_{\pm1}$
           & 88$_{\pm3}$ & 70$_{\pm2}$
           & 78$_{\pm2}$ & 73$_{\pm4}$ \\ \hline
\end{tabular}
\label{table:number_of_events}
\end{table}

The DEP, SEP and 2.6~MeV peaks are all
prominant in Figure~\ref{fig:energy_data_mc}a.
Only the SEP peak is also prominant
in Figure~\ref{fig:energy_data_mc}b.
The 511~keV annihilation photon that escapes the segmented detector
is fully absorbed by the REGe in these events.
The DEP~peak disappears because
the two 511~keV photons are emitted 
back to back and only one of the two photons 
can be tagged by the REGe detector.
The numbers of events in all samples are given in the
first row of Table~\ref{table:number_of_events}.

\begin{figure}[th!]
\center
\includegraphics[scale=0.5]{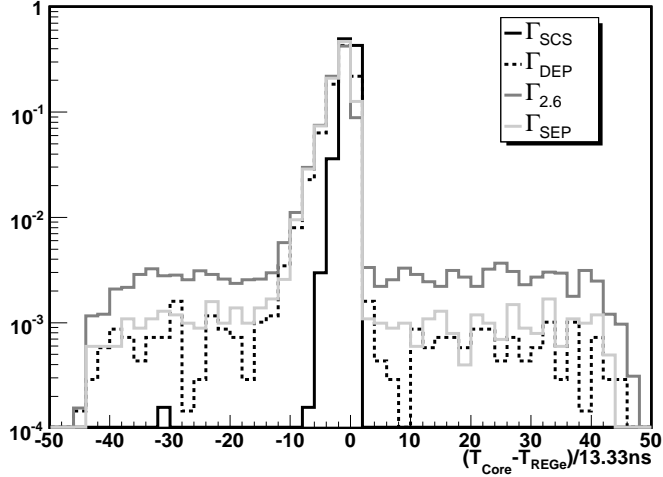}
\caption{$T_{Core}$-$T_{REGe}$ distributions.
The unit of the $x$ axis is the sampling clock.}
\label{fig:triggertime_diff}
\end{figure}

The time between the arrival of the core trigger ($T_{Core}$)
and the REGe trigger ($T_{REGe}$),
$\Delta T$ = $T_{Core}$-$T_{REGe}$, is shown
in Figure~\ref{fig:triggertime_diff}.
The $\Delta T$ distribution of the $\Gamma_{SCS}$ events
has a mean value of -9.4~ns with a RMS of 12.4~ns.
Only one event falls outside the Gaussian peak
($|\Delta T|$ more than 8$\times$13.3=107~ns).
This confirms that events in the $\Gamma_{SCS}$ sample
are predominantly induced by 2614~keV photons
from the $^{208}$Tl decay
and the fraction of accidental coincidences is negligible
at the 10$^{-4}$ level.

The $\Delta T$ distributions of
the $\Gamma_{2.6}$, $\Gamma_{DEP}$ and $\Gamma_{SEP}$ 
samples are also shown in 
Figure~\ref{fig:triggertime_diff}.
These $\Delta T$ distributions
are composed of ``signal'' peaks at $\Delta T$$\approx$0
and flat distributions of accidental coincidences.
The ``signal'' events in the $\Gamma_{DEP}$ and $\Gamma_{SEP}$ samples
register the 2.6~MeV photon in the segmented detector
through pair production
with one annihilation photon reaching the REGe detector.
The ``signal'' events in the $\Gamma_{2.6}$ sample
have another photon from the same $^{208}$Tl decay registered
in the REGe.
The numbers of accidental coincidence events can be 
calculated by fitting the $\Delta T$ distributions
with $|\Delta T|>80$~ns with a constant function.
The fractions of ``signal'' events
after subtracting the accidental coincidence events
are indicated by $f_{c}$ and
given in Table~\ref{table:number_of_events}.
The fractions of accidental coincidence events
(1-$f_{c}$)
agree with the rough estimate of $\approx$20\%
from the trigger rates, as explained
in Section~\ref{section:setup}.
Notice, that most accidental coincidence events
in the $\Gamma_{DEP}$, $\Gamma_{SEP}$ and $\Gamma_{2.6}$ samples
can be treated as events triggered with only the core of the
segmented detector and they
are actually classified correctly.
This was concluded in~\cite{cite:siegfried_dep_psa} where a
detailed study of core~only triggered events was presented.

The $\Gamma_{2.6}$, $\Gamma_{DEP}$ and $\Gamma_{SEP}$
samples have wider $\Delta T$ distributions
than the $\Gamma_{SCS}$ sample.
This is an artefact of the fixed 100~keV energy threshold applied
to the REGe detector. As the overall rise-time of a pulse,
see Figure~\ref{fig:pulse_example}a, does not depend on the energy,
the time at which a fixed threshold is reached does.
The $\Gamma_{2.6}$, $\Gamma_{DEP}$ and $\Gamma_{SEP}$
samples are selected 
without any cut on $E_{REGe}$. This results in 
much wider spreads in
$E_{REGe}$ and thus in wider
$\Delta T$ distributions.

\subsection{MC simulation}
\label{section:mc}

The GEANT4 based Monte Carlo package MaGe~\cite{cite:MaGe}
is used to simulate the setup.
In order to speed up the computation
only the $^{208}$Tl decay is simulated and not 
the complete decay chain of the $^{228}$Th source.
The energies as deposited in the germanium
detectors are smeared event by event according
to the detector resolutions.
The same energy thresholds and the coincidence
trigger as for the measured data are applied to the simulated events.
The MC is normalized to the data 
by counting the number of events within
the energy region of $E_{Core}+E_{REGe}$=2614$\pm$5~keV,
since events satisfying this requirement
are almost exclusively induced by the 2614~keV photon
from the $^{208}$Tl decay (see previous section).

The simulated distributions of $E_{Core}$ are shown in 
Figure~\ref{fig:energy_data_mc}a and b.
The same selection cuts as required for the 8 data samples
are applied to the MC events.
The data to MC ratios
are given in Table~\ref{table:number_of_events}.
They agree with the fractions of
events with true coincident triggers ($f_{c}$)
within $\approx$10\%.
The overall excess of data of $\approx$20\%
for all but the SCS samples
agrees well with the accidental coincidence rate.

\subsection{Distinction between MSE and SSE in MC}

\begin{figure}[th!]
\center
\includegraphics[scale=0.65]{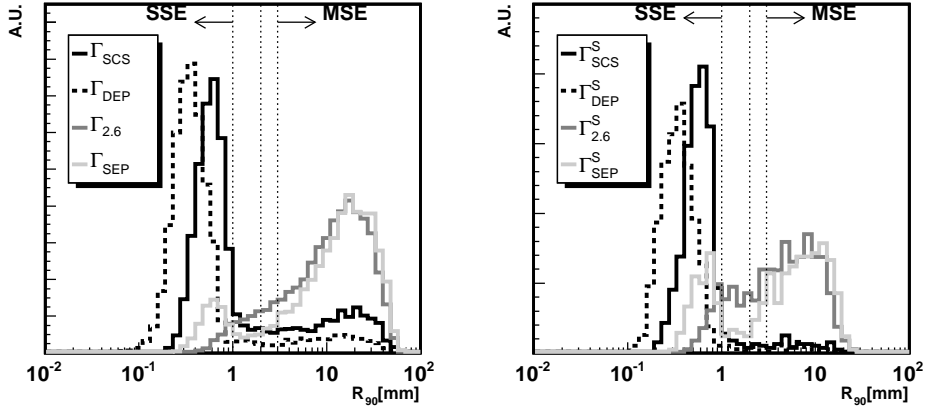}
\caption{
R$_{90}$ distributions of the 8 selected samples.
The 3 dotted vertical lines on each plot
indicate the $R_{90}$ values of 1, 2 and 3~mm.
}
\label{fig:r90}
\end{figure}

The variable $R_{90}$ is defined as
the radius of the volume that contains
90\% of the total energy deposition in a germanium detector.
It is used to study the size of the volume
within which the energy is distributed.
Details are described in~\cite{cite:siegfried_mc}.
The distributions of $R_{90}$ as calculated using MC information
are shown in Figure~\ref{fig:r90} for
the 8 selected samples.
Events from $\Gamma_{DEP}$ and $\Gamma_{SCS}$ samples
mostly have much smaller $R_{90}$ than those from
$\Gamma_{SEP}$ and $\Gamma_{2.6}$ samples.
$\Gamma_{SCS}$ events have slightly larger $R_{90}$ than
$\Gamma_{DEP}$ events
due to the higher energy of the recoiling electron.

A fraction of the SCS events
have relatively
large $R_{90}$ ($>$~2~mm).
In most of these events the 2.6~MeV 
photon Compton scatters several
times inside the segmented detector 
before reaching the REGe detector.
They still survive the $\Gamma_{SCS}$ cuts due to the relatively
large geometrical acceptance of the REGe detector.
Events with $R_{90}$$>$2~mm 
in the $\Gamma_{DEP}$ sample originate from
photons not interacting with the detector through pair production,
but through multiple Compton scattering, and still depositing
the same amount of energy as in DEP events.
These events are significantly reduced
by applying a single-segment cut, as shown in Figure~\ref{fig:r90}.
The fraction of events from the $\Gamma_{2.6}$
and $\Gamma_{SEP}$ samples with $R_{90}$$<$2~mm
have the high energy photon depositing energy very locally.
These fractions of events increase after applying
a single-segment cut.

The ``position resolution'' of
the DAQ is $\approx$~1~mm,
as explained in Section~\ref{section:setup}.
However, a conservative cut of $R_{90}$$<$2~mm is used to 
distinguish SSEs from MSEs~\cite{cite:siegfried_dep_psa}.
The fractions of SSEs ($f_{SSE}$) in each sample
are listed in Table~\ref{table:fraction_of_sse}.
The errors on $f_{SSE}$ are estimated by
varying the $R_{90}$ cut value between 1 and 3~mm.
\begin{table}[th!]
\center
\caption{Fractions $f_{SSE}$ of
         events with $R_{90}$$<$2$mm$ in each sample.
         }
\begin{tabular}{|c|c|c|c|c|}
\hline
sample     &  $\Gamma_{SCS}$ & $\Gamma_{2.6}$
           & $\Gamma_{DEP}$ & $\Gamma_{SEP}$ \\ \hline
$f_{SSE}$  &
   72 $^{+  3}_{-  6}$ \% & 
   10 $^{+  6}_{-  7}$ \% & 
   88 $^{+  1}_{-  2}$ \% & 
   15 $^{+  3}_{-  3}$ \% 
 \\ \hline \hline
sample     &  $\Gamma_{SCS}^{S}$ & $\Gamma_{2.6}^{S}$
           & $\Gamma_{DEP}^{S}$ & $\Gamma_{SEP}^{S}$ \\ \hline
$f_{SSE}$  &
   92 $^{+  1}_{-  3}$ \% & 
   26 $^{+  12}_{- 15}$ \% & 
   96 $^{+  1}_{-  1}$ \% & 
   31 $^{+  6}_{-  5}$ \% 
 \\ \hline
\end{tabular}
\label{table:fraction_of_sse}
\end{table}
$\Gamma_{SCS}$ has a smaller fraction of SSEs
than $\Gamma_{DEP}$,
due to the relatively large selection window.
The $f_{SSE}$ fractions for the $\Gamma^S$ samples are larger
than for the $\Gamma$ samples,  since the single-segment cut
already removes most MSE events.

If only the segmented detector is used for triggering,
$f_{SSE}$~$\approx$78\% for the $\Gamma_{DEP}$ sample,
and $\approx$12\% for the $\Gamma_{2.6}$ sample~\cite{cite:siegfried_dep_psa} 
(89~\% and 30\% for $\Gamma_{DEP}^S$ and 
$\Gamma_{2.6}^S$ samples, respectively).
These values are similar to the ones for coincident events.
Therefore, even though accidental coincidences are not 
simulated by the MC, the $f_{SSE}$ values as presented
in Table~\ref{table:fraction_of_sse} can be used to
evaluate the data samples 
collected with the coincidence trigger.

If the estimated 1~mm position resolution 
can be achieved through PSA,
the SSEs from each sample should be correctly identified.
The PSA procedure is described in the following section.

\section{Pulse shape analysis}
\label{chapter:psa}

The same Artificial Neural Network (ANN) package
as used in~\cite{cite:siegfried_dep_psa}
is used here to perform the pulse shape analysis.
The ANN is trained with
an SSE sample against an MSE sample.
In~\cite{cite:siegfried_dep_psa}
$\Gamma_{DEP}$ (without coincidence trigger)
was used as the SSE--dominant sample
and events in the 1620~keV line
(with the 1620~keV photon from $^{212}$Bi decay
fully absorbed in the segmented detector)
as the MSE--dominant sample.
The trained ANN was able to identify
both SSE and MSE events with $\approx$~85\% efficiencies.

In this study, a similar analysis is performed.
The ANN is trained
with the $\Gamma_{DEP}$ sample (SSE--dominant)
against the $\Gamma_{SEP}$ sample (MSE--dominant).
The trained ANN is used to
verify that the collected events 
in the $\Gamma_{SCS}$ sample are SSE--dominant.
The results are shown in section~\ref{sec:psa_verification}
after a general description in section~\ref{sec:ann_and_r90}.

In a second analysis the ANN is trained with
the $\Gamma_{SCS}$ against the $\Gamma_{2.6}$ sample.
It is shown in section~\ref{sec:psa_results}
that the results are consistent.

\subsection{General features of the ANN}
\label{sec:ann_and_r90}

\begin{figure}[th!]
\center
\includegraphics[scale=0.65]{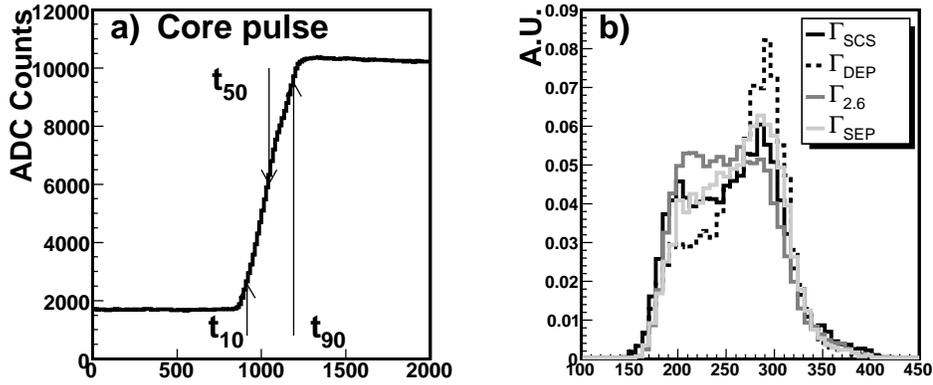}
\caption{a) Core pulse of one DEP event,
$t_{10}$, $t_{50}$ and $t_{90}$
are indicated by arrows.
b) distributions of the rise time $T_r$=$t_{90}-t_{10}$
for the 4 samples under consideration.}
\label{fig:pulse_example}
\end{figure}

The core pulse of the segmented detector
of a typical $\Gamma_{DEP}$ event
is shown in Figure~\ref{fig:pulse_example}a.
The rising part of the pulse contains information
about the event structure
as explained in Section~\ref{section:introduction}.
The time $t_{50}$ is
defined as the time at which the pulse has reached
50\% of its maximum\footnote{
Pedestals are subtracted by using the information
during the 1~$\mu$s interval before the trigger. 
}.
The 20~values before and the~20
after~$t_{50}$ are used for PSA.
Thus, the selection of the 40 values is 
independent of the absolute amplitude of the pulse and 
thus independent of the energy.

$t_{10}$ and $t_{90}$ are defined as
the times when the pulse
reaches 10\% and 90\% of its maximum, respectively.
The distributions of the pulse rise time, $T_{r}$=$t_{90}$-$t_{10}$,
are shown in Figure~\ref{fig:pulse_example}b.
$T_r$ is fully covered by the 40 values
which cover a time window of 533~$ns$.
The dominance of long risetimes in the $\Gamma_{DEP}$
sample reflects the dominance of events close to the
detector surface.

The ANN package 
as used here
has 40 input neurons for 
the 40 pulse values.
It has two hidden layers with 8 and 2 neurons each
and 1 output neuron.
The ANN is trained such that 
a large ANN output ($NN_{out}$) indicates
that the event is SSE--like
and a small $NN_{out}$ indicates that it is MSE--like.

Since both $NN_{out}$ and $R_{90}$ are related
to the size of the energy deposition in the detector,
a correlation between
$R_{90}$ and $NN_{out}$ is predicted.
On average events with small $R_{90}$ should have
large $NN_{out}$ and vice versa.
It is clear that $R_{90}$ is not the 
only variable that determines
the pulse shape. 
Other, second order effects like the drift anisotropies caused
by the crystal structure and inhomogenious doping concentrations
also modify the pulse shapes.
Therefore, a 100\% correlation between
$NN_{out}$ and $R_{90}$ is not expected.
The details of this correlation can only be studied with
a detailed pulse shape simulation which
is beyond the scope of this paper.

\subsection{Verification of ANN training with single Compton scattering
events}
\label{sec:psa_verification}

\begin{figure}[th!]
\center
\includegraphics[scale=0.75]{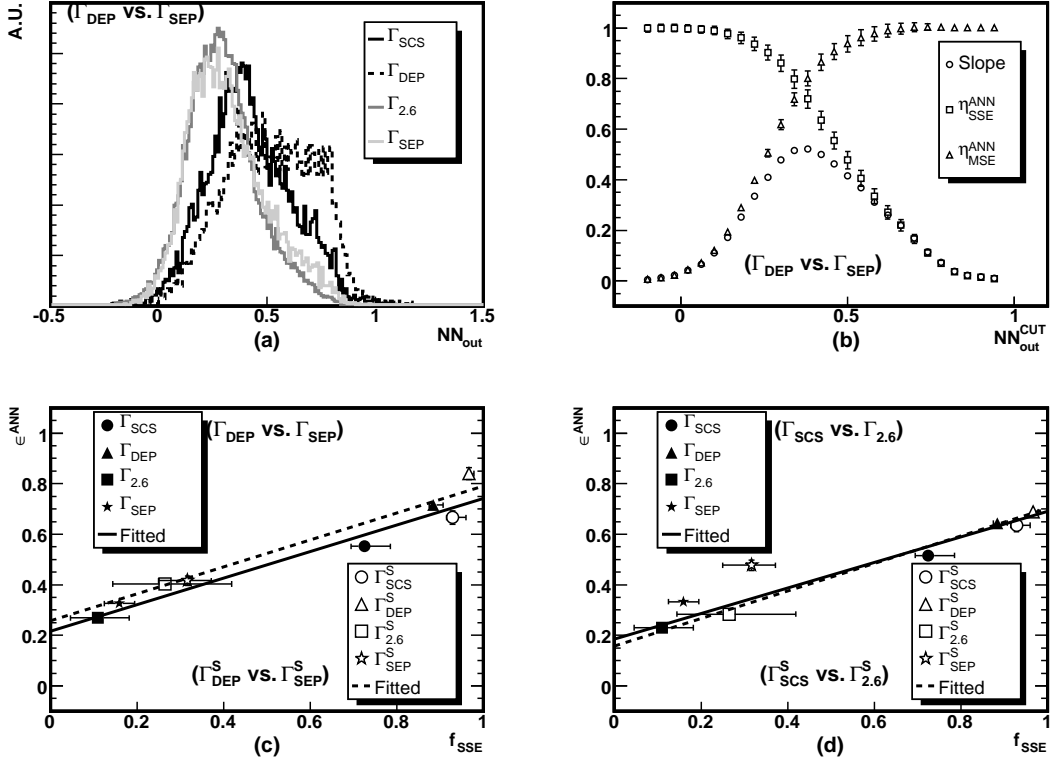}
\caption{
a-c) Results of the ANN analysis, if the ANN is trained 
with the $\Gamma_{DEP}$ sample as SSE--dominant 
and the $\Gamma_{SEP}$ sample as  MSE--dominant:
a) $NN_{out}$ distributions for all four samples.
b) $\eta^{ANN}_{SSE}$ and $\eta^{ANN}_{MSE}$ vs. $NN_{out}^{CUT}$.
The fitted slope $a$, see text, is shown as well.
Errors are taken from the MINUIT fit.
c) $\epsilon^{ANN}$ vs. $f_{SSE}$;
$\epsilon^{ANN}$ 
values correspond to the value of $NN_{out}^{CUT}$ 
giving the maximum fitted slope $a$.
Also given are results for the single segment samples indicated
by $^S$~(open points). 
d) 
Cross--check with the ANN trained 
with the $\Gamma_{SCS}$ sample as SSE--dominant 
and the $\Gamma_{2.6}$ sample as  MSE--dominant: 
$\epsilon^{ANN}$ vs. $f_{SSE}$. Results are again
also given for single segment events. 
See text for details.
}
\label{fig:psa_dep_vs_sep}
\end{figure}

The ANN is trained with the $\Gamma_{DEP}$ sample
as SSE--dominant (signal--like)
and the $\Gamma_{SEP}$ sample as MSE--dominant (background--like). 
The training takes 300 iterations.\footnote{
The ANN trained with 500 iterations gives similar results.}
The trained ANN is then applied to
all $\Gamma_{SCS}$ and $\Gamma_{2.6}$ events.
It should correctly identify them as single--site and the
multi--site events.
The $NN_{out}$ distributions for all 4 samples are shown 
in Figure~\ref{fig:psa_dep_vs_sep}a.

The $\Gamma_{SCS}$~events have in average larger $NN_{out}$ values
than the $\Gamma_{2.6}$ events. The peaks of the distributions
are well separated.
However, while the distribution for $\Gamma_{2.6}$~events is
quite similar to the one for $\Gamma_{SEP}$~events,
the distribution for the $\Gamma_{SCS}$~events looks different
from the one for $\Gamma_{DEP}$~events.
A shift of the peak is expected from the MC simulation, since there is
a higher percentage of
$\Gamma_{SCS}$~events 
with $R_{90}$~values above 2~mm indicating an MSE--like structure
of the events, see Figure~\ref{fig:r90}. 
The $\Gamma_{DEP}$ distribution in addition features
a plateau towards high $NN_{out}$ values.
This is probably an artefact of the spatial distribution of
the events which are predominantly close to the surface
which also influenced the ANN training.

The classification of events using the distributions 
depicted in Figure~\ref{fig:psa_dep_vs_sep}a
is based on a cut in $NN_{out}$, $NN_{out}^{CUT}$.
An event is classified as 
SSE--like, if $NN_{out}$$>$ $NN_{out}^{CUT}$,
or MSE--like, if $NN_{out}$$<$$NN_{out}^{CUT}$.
For a given value of $NN_{out}^{CUT}$,
the survival efficiency for any data sample,
$\epsilon^{ANN}$, is defined as the fraction of
events in that sample that are
identified by the ANN as SSE--like events.

The probabilities to correctly identify 
SSE-- and MSE--like events,
$\eta^{ANN}_{SSE}$ and $\eta^{ANN}_{MSE}$,
are calculated using the Monte Carlo predictions for the
purities $f_{SSE}$ of the samples used, 
see Table~\ref{table:fraction_of_sse},
and using the measured $\epsilon^{ANN}$ for the data samples.
A linear dependence 
$\epsilon^{ANN}=a\times f_{SSE}+b$
is assumed. 
For a given $NN_{out}^{CUT}$, 
the values for  $\epsilon^{ANN}$ are calculated for all samples,
a linear fit is performed to obtain the slope 
and the line is extrapolated to 
$f_{SSE}$=1 to obtain 
$\eta^{ANN}_{SSE}$. It is extrapolated to
$f_{SSE}=0$ to determine 1-$\eta^{ANN}_{MSE}$
(see Figures~\ref{fig:psa_dep_vs_sep}c and d for two fits).
The fit procedure takes errors into account.
The errors on $f_{SSE}$ are
listed in Table~\ref{table:fraction_of_sse}
and those on $\epsilon^{ANN}$ are
statistical only.
The resulting $\eta^{ANN}_{SSE}$ and $\eta^{ANN}_{MSE}$
as a function of $NN_{out}^{CUT}$ are shown in
Figure~\ref{fig:psa_dep_vs_sep}b.
The fitted slope~$a$ is shown in Figure~\ref{fig:psa_dep_vs_sep}b
as a function of $NN_{out}^{CUT}$ as well.
A clear maximum for $a$ is visible.

The correlations between the values of $\epsilon^{ANN}$ 
and $f_{SSE}$  are shown in
Figure~\ref{fig:psa_dep_vs_sep}c
for the value of $NN_{out}^{CUT}$ which maximizes the
slope $a$.
The slope $a$ 
does not approach the ideal value of~1,
indicating that $f_{SSE}$ and $NN_{out}$ are not fully correlated.
This is expected as the predictions for $f_{SSE}$ 
are entirely based on the simple variable~$R_{90}$ 
as discussed in Section~\ref{sec:ann_and_r90}.
The results for the single segment samples are also shown.
They were subjected to the identical analysis using
the equivalent samples for training.
The results of the fits are indicated for both single segment and
unrestricted event samples.

The results for $\eta^{ANN}_{SSE}$ and $\eta^{ANN}_{MSE}$
are given in the first two rows of Table~\ref{table:psa_results}
with errors deduced from the fits.
The ANN can correctly
identify both SSE and MSE events at the 75~\% to 80~\% level.
The results for the single segment data sets
are similar to ones for the unrestricted samples.
These results agree in general with the values of $\approx$85~\%
as achieved in~\cite{cite:siegfried_dep_psa}.

The compatability of the points with the linear fits in
Figure~\ref{fig:psa_dep_vs_sep}c
leads to the conclusion that the
SSE--like events in the $\Gamma_{SCS}$~sample are identified
with about the same efficiency as in the other samples.
This is the most important result of this study indicating
that tagged SCS~events can indeed be used to further study
pulse-shapes in more detail.

\begin{table}[th!]
\center
\begin{tabular}{|c|c||c|c|}
\hline
\multicolumn{2}{|c||}{ANN Training} & \multicolumn{2}{c|}{Analysis} \\ \hline\hline
SSE-dominant & MSE-dominant & $\eta_{SSE}^{ANN}$ &  $\eta_{MSE}^{ANN}$ \\ \hline
$\Gamma_{DEP}$ &  $\Gamma_{SEP}$ & $74.1_{\pm 2.7}$ \% & $78.3_{\pm 2.8}$ \% \\ \hline
$\Gamma_{DEP}^S$ & $\Gamma_{SEP}^S$ & $79.1_{\pm 7.2}$ \% & $74.3_{\pm 6.8}$ \% \\ \hline
$\Gamma_{SCS}$ & $\Gamma_{2.6}$ & $69.0_{\pm 2.1}$ \% & $81.5_{\pm 2.5}$  \% \\ \hline
$\Gamma_{SCS}^S$ & $\Gamma_{2.6}^S$ & $70.2_{\pm 4.3}$ \% & $84.2_{\pm 5.1}$  \% \\ \hline
\end{tabular}
\caption{$\eta^{ANN}_{SSE}$ and $\eta^{ANN}_{MSE}$ 
with the ANN trained with various SSE-dominant samples
against various MSE-dominant samples.}
\label{table:psa_results}
\end{table}

\subsection{Cross-check using SCS~events for ANN training} 
\label{sec:psa_results}

The same procedure as described in the previous section is repeated with
the ANN trained using the $\Gamma_{SCS}$ ($\Gamma_{SCS}^S$) 
as the SSE-dominant
and the $\Gamma_{2.6}$ ($\Gamma_{2.6}^S$) as the MSE-dominant samples.
The values of $\epsilon^{ANN}$ versus $f_{SSE}$ corresponding
to the maximum slope are shown 
in Figure~\ref{fig:psa_dep_vs_sep}d.

The resulting identification probabilities
$\eta^{ANN}_{SSE}$ and $\eta^{ANN}_{MSE}$ are given
in the last two rows of Table~\ref{table:psa_results}.
The ANN can correctly
identify SSE--like events at the 70~\% and MSE--like events 
at the 80~\% level.
This confirms again that the selected
SCS~samples are enriched in SSE--like events and
can be used to train the ANN package.

\section{Conclusions and outlook}
\label{chapter:conclusion}

Events with photons Compton scattering only once inside
a germanium detector, SCS events, 
can be selected by tagging the scattered photon with
a second germanium detector.
The pulse shapes of these events can be studied and used to
test methods that distinguish between
single-site and multi-site events.

In order to collect SCS events
and perform pulse shape analysis,
an 18-fold segmented prototype detector
for the Phase-II of the GERDA experiment
was positioned in front of a $^{228}$Th source.
A second germanium detector was positioned
to record the escaped photons at 72$\textdegree$,
corresponding to 2040~keV energy deposit
in the segmented detector, close to
the $Q$-value of the 0$\nu$$\beta\beta$ decay of $^{76}$Ge.

According to the MC simulation
$\approx$72~\% of the collected SCS events 
are true SSE events.
The SSE-dominance is verified 
by an artifical neural network 
(ANN) trained in an independent way.
These SCS events are then themselves used to train the pulse shape analysis
package and thus the trained PSA is able to identify single-
and multi-site events with efficiencies at the $\approx$80\% level.

Future studies can improve in two ways.
The fraction of SSE events in the collected SCS sample
can be increased by further improving the tagging method.
For example, the whole experimental setup can be shielded
from external photons and
collimators can be positioned between the two detectors.
The Monte Carlo predictions can also be improved.
Currently they are based on the size of the energy
deposits only.
Better predictions 
require a detailed pulse shape simulation
which is currently being developed for the detectors under study.


\addcontentsline{toc}{section}{Bibliography}


\begin{thebibliography}{99}
%
\bibitem{cite:siegfried_dep_psa}
I.~Abt {\it et al.}, arXiv:0704.3016,
submitted to EJC, to be published.

\bibitem{cite:gerda_white_paper}
S. Sch\"onert {\it et al.} [GERDA Collaboration],
Nucl. Phys. Proc. Suppl. {\bf 145} (2005) 242.

\bibitem{cite:first_scs}
F. Petry, {\it et al.},Nucl. Instr. and Meth.
{\bf A 332} (1993) 107.

\bibitem{cite:HdMo}
J. Hellmig and H.V. Klapdor-Kleingrothaus,
Nucl. Instr. and Meth.
{\bf A 455} (2000) 638.

\bibitem{cite:HdMo_dep}
B. Majorovits and H.V. Klapdor-Kleingrothaus,
Eur.Phys. J. {\bf A 6} (1999) 463.

\bibitem{cite:igex}
D. Gonz\~alez, {\it et al.}, Nucl. Instr. and Meth.
{\bf A 515} (2003) 634.

\bibitem{cite:dep_part_majorana}
S.R. Elliott, V.M. Gehman, K. Kazkaz, D-M. Mei, A.R. Yong,
Nucl. Instr. and Meth. {\bf A 558} (2006) 504.

\bibitem{cite:siegfried_mc}
I.~Abt {\it et al.}, Nucl. Instr. and Meth.
{\bf A 570} (2007) 479-486.

\bibitem{cite:pixie4_daq}
User's Manual Digital Gamma Finder (DGF) PIXIE-4,
XIA LLC,  {\it http://www.xia.com}


\bibitem{cite:siegfried_characterization}
I.~Abt {\it et al.}, Nucl. Instr. and Meth. {\bf A 577} (2007) 574


\bibitem{cite:siegfried_photon_identification}
I.~Abt {\it et al.}, nucl-ex/0701005, submitted to
NIM, to be published.


\bibitem{cite:rege}
Canberra Reverse-Electrode Coaxial Ge Detector,\\
{\it http://www.canberra.com/Products/494.asp}

\bibitem{cite:MaGe}
M.~Bauer {\it et al.},
Journal of Physics, Conf. Series. {\bf 39} (2006) 362.

%
\end{thebibliography}
\end{document}